\documentclass[11pt]{article}
\usepackage[T1]{fontenc}
\usepackage[utf8]{inputenc}
\usepackage{geometry}
\usepackage{float}
\usepackage{textcomp}
\usepackage{amsmath}
\usepackage{graphicx}
\usepackage{subfig}
\usepackage{setspace}
\usepackage{microtype} 
\usepackage{color}
\usepackage[symbol]{footmisc}
\usepackage{titlesec}

\makeatletter

\newcommand{\lyxmathsym}[1]{\ifmmode\begingroup\def\b@ld{bold}
	\text{\ifx\math@version\b@ld\bfseries\fi#1}\endgroup\else#1\fi}

\usepackage{slashed}
\usepackage[colorlinks=true,breaklinks=true]{hyperref}

\usepackage[numbers,sort&compress]{natbib}

\makeatother

\usepackage{babel}

\begin{document}
	
\begin{titlepage}
\title{{\bf{Study of rare top quark decays into a jet plus a charged pseudo-scalar meson }}}
\date{}
\maketitle
\vspace{-15mm}
\thispagestyle{empty}
\maketitle
\begin{center}       
\author{\bf{Long-Shun Lu$^{a,b}$, 
Lei-Yi Li$^{a,b,c}$\footnote{Corresponding author: lileiyi@sjtu.edu.cn}, 
Cai-Dian L\"u$^{a,b}$

\vskip0.5cm         
{ \it $^a$ Institute of High Energy Physics, CAS, P.O. Box 918(4) Beijing 100049,  China} \\         
{ \it $^b$ School of Physics, University of Chinese Academy of Sciences, Beijing 100049, China} \\ 
{ \it $^c$ INPAC, Key Laboratory for Particle Astrophysics and Cosmology (MOE), \\
Shanghai Key Laboratory for Particle Physics and Cosmology, \\
School of Physics and Astronomy, Shanghai Jiao Tong University, Shanghai 200240, China}}  }     
\end{center}
\vskip1cm
\begin{abstract}
The semi-inclusive decay processes of a top quark into a
charged pseudo-scalar meson and a jet are studied within
the framework of QCD factorization. The leading power of the decay matrix elements can be factorized into heavy-to-light quark transition current and a
hadron matrix element up to next-to-leading order QCD corrections. We calculate one-loop virtual corrections together with real gluon emission corrections at the $\alpha_s$ order. The numerical results of the branching ratios are presented for the sum of two-body and three-body decays. We also study the energy cutoff dependence of the gluon jet.  These processes are hopeful to be detected in the near future experiments, which can serve as probes for new physics.

\newpage{}
\end{abstract}
\end{titlepage}

\section{Introduction}

As the most massive particle in the Standard Model, top quark is believed to be most sensitive to new physics
beyond the standard model. The study of top quark decay is one of the most hot topics in both theoretical and
experimental studies.  Due to its
very short lifetime, the top quark decays before hadronization, making
it unique among all quarks. Many studies have focused on calculating
the dominant decay process $t\rightarrow bW^{+}$
\cite{Jezabek:1988iv,Czarnecki:1990kv,Li:1990qf,Gao:2012ja,Chen:2023dsi},
which allows for the determination of intrinsic properties such as
the top quark's mass and lifetime. Recently, there
has been growing interest in rare top quark decays \cite{Handoko:1998tz,Handoko:1998uc,Beneke:2000hk,dEnterria:2023wjq,Handoko:1999iu,dEnterria:2020ygk,Liu:2004qw,Cai:2022xha,Cai:2024xjq},
as these processes are sensitive to new physics and can serve as probes
to study beyond the Standard Model \cite{Liu:2004qw,Cai:2022xha,Cai:2024xjq}.
Among these decays, a particularly interesting type of rare decays is the top
quark semi-inclusive decay \cite{Beneke:2000hk,dEnterria:2023wjq,Handoko:1999iu,dEnterria:2020ygk}, where the top quark decays into a single high-energy
meson and a jet. Theoretically, these processes can be computed perturbatively,
while experimentally, different decay channels can be identified by the distinct final states of the meson and jet, enabling their measurement.

There are two distinct types of  the semi-inclusive decay channels of the top quark: one involves a final state containing an down-type quark jet and a charged meson 
 $t\rightarrow qM^{+}$ \cite{Beneke:2000hk,dEnterria:2023wjq}, while the other involves a final state containing a up-type quark jet and a neutral meson $t\rightarrow qM^{0}$ \cite{dEnterria:2023wjq,Handoko:1999iu,dEnterria:2020ygk}.
In Ref.~\cite{Beneke:2000hk,dEnterria:2023wjq}, the leading order (LO) results
for the top quark semi-inclusive decay processes $t\rightarrow b M^{+}$ are presented. Ref.~\cite{Handoko:1999iu}
calculates the branching ratio for the CKM-suppressed process $t\rightarrow c\Upsilon$
using Non-relativistic QCD (NRQCD), and discusses it in the context
of the Minimal Supersymmetric Standard Model. Ref.~\cite{dEnterria:2020ygk}
employs the NRQCD method to calculate the decay of a top quark into
a neutral meson and an up-type quark jet, with predictions for potential
observation processes $t\rightarrow\bar{B}^{0}+\mathrm{jet}$ and
$t\rightarrow\bar{B}_{s}+\mathrm{jet}$ in future experiments.

In this work, we study the top quark semi-inclusive decay $t\rightarrow qP^{+}$, where
$q=d,s,b$ and $P^{+}$ denotes a pseudo-scalar meson. The final states
of these semi-inclusive decays include mesons, making them suitable
for calculation within the framework of QCD factorization \cite{Beneke:1999br,Beneke:2001ev,Cheng:2001nj,Beneke:2003zv,Grossman:2015cak,Alte:2015dpo,Lu:2022kos}. The leading order decay amplitude for $t\rightarrow qP^{+}$ processes
can be easily factorized into the decay constant of the final state meson and the heavy-to-light quark current. Given the heavy mass of the top quark, the subleading
power ($1/m_t$) corrections to the $t\rightarrow bP^{+}$ process are small,
 therefore the next-to-leading-order corrections primarily arise from the QCD loop
corrections.

In our  study, we calculate the next-to-leading order (NLO) QCD corrections
to the $t\rightarrow qP^{+}$ processes, where
the ultraviolet divergences in the loop diagrams are renormalized using
dimensional regularization. For the infrared divergences from virtual corrections in  these processes, we eliminate them by
accounting the real gluon emission. The numerical results indicate
that the branching ratios for these processes span the range of $10^{-16}\sim10^{-7}$.
The processes $t\rightarrow b\pi^{+}$ and $t\rightarrow bD_{s}^{+}$
are expected to be measurable at future colliders, such as the High-Luminosity
Large Hadron Collider (HL-LHC) \cite{Apollinari:2015wtw}, Super Proton-Proton
Collider (SPPC) \cite{thecepcstudygroup2018cepc,Tang:2022fzs} and
Future Circular Hadron Collider (FCC-hh) \cite{FCC:2018vvp}. Some of the CKM-suppressed rare
processes are computed {for the first time}, which could serve as probes
for new physics.

The structure of the paper is as follows: the theoretical framework,
including the calculation of the one-loop diagrams, is provided in
the next section. In the third section, we present the numerical results
and a phenomenological discussion, followed by the conclusions in the
final section.

\section{Theoretical framework}


Both the mass of the final-state
pseudo-scalar meson $m_{P}$ and the mass of the final-state jet $m_{q}$
are much smaller than the mass of the top quark $m_{t}$. Therefore,
we can safely neglect the final state masses and assume that they lie on the
light-cone. The light-cone vectors are defined as
\begin{equation}
n=(1,0,0,1),\qquad\bar{n}=(1,0,0,-1),
\end{equation}
which  satisfy the relations $n^{2}=\bar{n}^{2}=0$
and $n\cdot\bar{n}=2$. In the light-cone coordinate system, the momentum
of the final-state meson is denoted as $p$, aligned along the light-cone
direction, while the momentum of the quark jet is denoted as $k$,
aligned along the anti-collinear direction:
\begin{equation}
p^{\mu}=\bar{n}\cdot p\dfrac{n^{\mu}}{2}+n\cdot p\dfrac{\bar{n}^{\mu}}{2}+p_{\perp}^{\mu},\qquad k^{\mu}=\bar{n}\cdot k\dfrac{n^{\mu}}{2}+n\cdot k\dfrac{\bar{n}^{\mu}}{2}+k_{\perp}^{\mu}.
\end{equation}

\begin{figure}[htb]
\centering{}\includegraphics[scale=0.46]{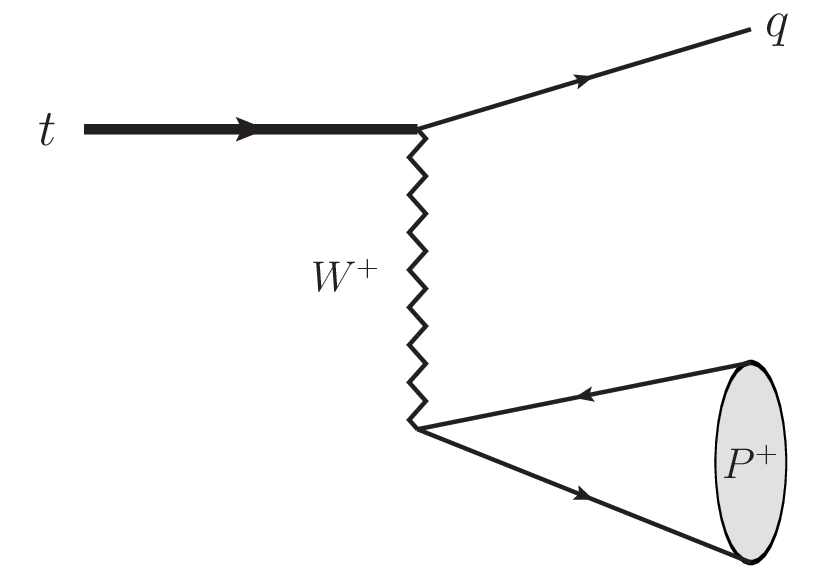}
\caption{Leading order Feynman diagram for the $t\rightarrow qP^{+}$ semi-inclusive decay.\label{fig:semi-inclusive-of-local-1}}
\end{figure}

In the leading power approximation, $\bar{n}\cdot p=n\cdot k=m_{t}$ represents
the large momentum component, while the other components are power suppressed. The momentum of the top quark is defined
as:
\begin{equation}
p_{t}^{\mu}=m_{t}v^{\mu},
\end{equation}
where $v^{\mu}=(1,0,0,0)$ represents the velocity of the top quark at its rest frame.
The tree-level diagram for the semi-inclusive decay process $t\rightarrow qP^{+}$
is shown in Fig.~\ref{fig:semi-inclusive-of-local-1} and the decay amplitude is 
\begin{equation}
    \begin{aligned}
    \mathcal{A}^{(0)} & =-\dfrac{4G_{F}}{\sqrt{2}}V_{q_{1}q_{2}}V_{tq}^{*}\langle P^{+}(p)\big|\bar{q}_{1}\gamma^{\mu}P_{L}q_{2} \big|0\rangle \langle  q(k)\big|\bar{q} \gamma_{\mu}P_{L}t  \big|t(p_{t}) \rangle,
    \end{aligned}
    \label{eq:tree-level factorization}
\end{equation}
where $V_{tq}^{*}$ and $V_{q_{1}q_{2}}$ represent the CKM matrix elements,
with the up-type quark $q_{1}=u,c$ and the down-type quark $q,q_{2}=d,s,b$. 
The $q$-jet is generated from the heavy-to-light current of the top quark, while the final-state pseudo-scalar meson
is produced from the $W$-boson. 
The definition of the decay constant
for the pseudo-scalar meson is given by:
\begin{equation}
\langle P^{+}(p)|\bar{q}_{1}\gamma_{\mu}\gamma_{5}q_{2}|0\rangle=-if_{P}p_{\mu},
\label{eq:decay constants}
\end{equation}
where $f_{P}$ represents the decay constant of the pseudo-scalar meson. In the QCD factorization
framework, the tree-level decay amplitudes of $t\rightarrow qP^{+}$
processes are proportional to the meson decay constant times the quark weak transition 
current. We then obtain the tree-level decay width of top quark\,\footnote{Our formula differs by a factor of 1/9 from that in Ref.~\cite{Beneke:2000hk}, while agreeing with the updated version of Ref.~\cite{dEnterria:2023wjq}.}
:
\begin{equation}
\Gamma^{(0)}=\dfrac{G_{F}^{2}m_{t}^{3}f_{P}^{2}}{16\pi}|V_{q_{1}q_{2}}V_{tq}^{*}|^{2}.
\label{eq:decay_width_LO}
\end{equation}

\subsection{Virtual corrections}\label{section2}

 As probes for studying
new physics, 
the experimental precision of top quark decay processes in future colliders \cite{Apollinari:2015wtw,thecepcstudygroup2018cepc,Tang:2022fzs,FCC:2018vvp} requires   necessitating precise calculations of the
processes to improve theoretical accuracy.  As discussed in the introduction, the
power correction is suppressed by $1/m_{t}$, leading to small corrections, since the mass of the top quark
is much larger than the masses of the final-state jet and meson.
The QCD loop contributions constitute the main contribution to the next-to-leading-order corrections.

\begin{figure}[htb]
\centering{}\includegraphics[scale=0.32]{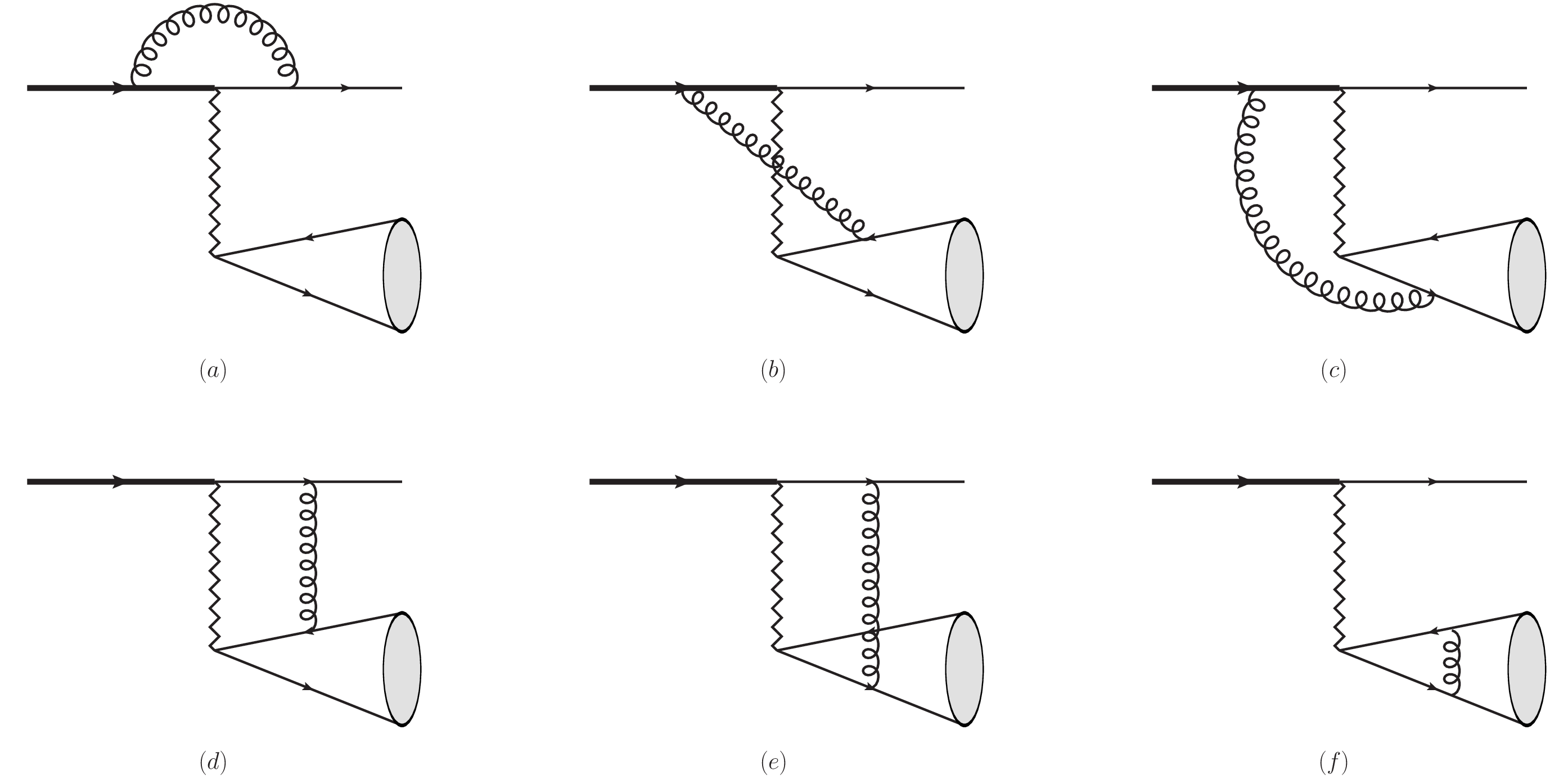}
\caption{Virtual correction for semi-inclusive processes of $t\rightarrow qP^{+}$ \label{fig:virtual}}
\end{figure}

The Feynman diagrams for the NLO QCD corrections to the processes \( t \rightarrow qP^{+} \) are shown in Fig.~\ref{fig:virtual}.
In diagrams (b) to (e), the colored gluon is coupled to the final-state meson which is a color singlet state; thus, the
contributions from these diagrams are zero. Diagram (f) can be absorbed into the definition of decay constants in Eq.~(\ref{eq:decay constants}). Therefore,
for the process $t\rightarrow qP^{+}$, it suffices to consider only
diagram (a), and the one-loop result for the decay amplitude still retains the factorization formalism as the tree-level in Eq.~(\ref{eq:tree-level factorization}). To address the ultraviolet divergences at the amplitude level, we employ dimensional regularization for subtraction, working in $D = 4- 2\epsilon$ dimension. Under the conditions
of on-shell renormalization, the wave function of the top quark is
renormalized as follows:
\begin{equation}
{Z_{t}=1-\dfrac{\alpha_{s}(\mu)}{4\pi}C_{F}\bigg(\dfrac{4\pi\mu^{2}}{m_{t}^{2}}\bigg)^{\epsilon}\bigg(\dfrac{3}{\epsilon}+4-3\gamma\bigg).}
\end{equation}
where  $\gamma$ is the Euler's constant. For the quark jet, we neglect its mass, resulting in renormalization
constant $Z_{q}=1$. Taking the renormalization into account,  the ultraviolet divergences in our loop diagram calculation can been removed in the standard way.

For the infrared divergence, we also work in the dimensional regularization. The
 decay width {in $D$-dimension} for the NLO QCD correction can be expressed as:
\begin{equation}
\Gamma_{\mathrm{virtual}}^{(1)}=-\dfrac{\Gamma^{(0)}}{m_{t}^{2\epsilon}}\bigg[\dfrac{\alpha_{s}(\mu)}{3\pi}\bigg(\dfrac{4\pi\mu}{m_{t}}\bigg)^{2\epsilon}\bigg(\frac{2}{\epsilon^{2}}+\frac{9-4\gamma}{\epsilon}+4\gamma^{2}-18\gamma-\frac{\pi^{2}}{3}+30\bigg)\bigg],
\label{eq:decay_width_virtual}
\end{equation}
%
where $1/m_{t}^{2\epsilon}$ is from the two-body phase space integral in dimensions $D$ \cite{Schwartz:2014sze}.  The remaining divergent terms in Eq.~(\ref{eq:decay_width_virtual})
correspond to infrared divergences, which require additional subtraction from real gluon emission contributions.

\subsection{Real gluon correction}

\begin{figure}
\centering{}\includegraphics[scale=0.4]{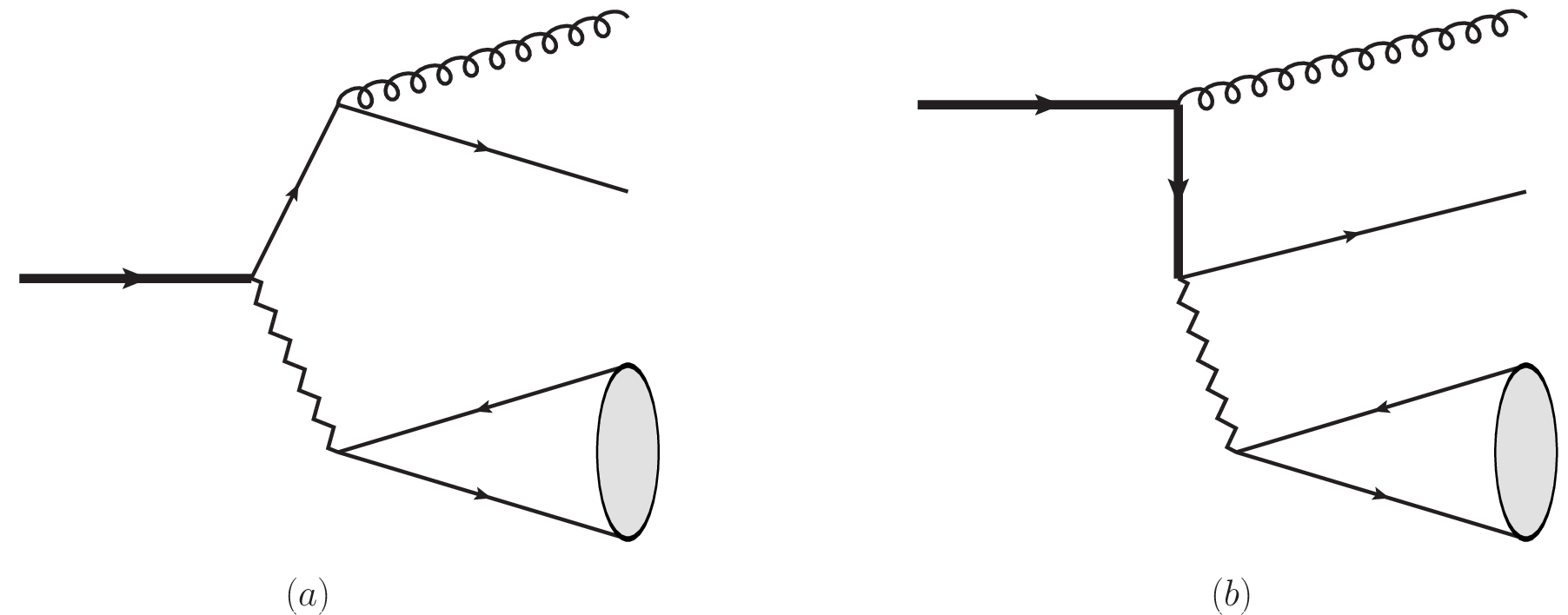}
\caption{Real gluon emission for the semi-inclusive processes of $t\rightarrow qP^{+}$ \label{fig:real}}
\end{figure}

As a semi-inclusive process, $t\rightarrow qP^{+}$ involves a heavy-to-light
current $t\rightarrow q$ that exhibits a single-jet effect. The radiation
of real gluons contributes to the decay width at order $\mathcal{O}(\alpha_{s})$, the same order as the virtual correction, to be considered together. In the calculations of Section
\ref{section2}, we neglected the mass of the bottom quark. Under the leading order of $1/m_t$ power expansion,
this introduces soft and collinear divergences, necessitating the
introduction of real corrections for subtraction. The Feynman diagrams
for real gluon radiation in the $t\rightarrow qP^{+}$ process are
shown in Fig.~\ref{fig:real}. The decay width of the three body decay resulting from the real
gluon radiation in $D$-dimensions is given by:
\begin{equation}
    \Gamma_{\mathrm{real}}^{(1)}=\dfrac{\Gamma^{(0)}}{m_{t}^{2\epsilon}}\bigg[\dfrac{\alpha_{s}(\mu)}{3\pi}\bigg(\dfrac{4\pi\mu}{m_{t}}\bigg)^{2\epsilon}\bigg(\frac{2}{\epsilon^{2}}-\frac{4\gamma-9}{\epsilon}-18\gamma+4\gamma^{2}-\dfrac{5}{3}\pi^{2}+35\bigg)\bigg],\label{3-body-decay}
\end{equation}
%
{where $1/m_{t}^{2\epsilon}$ is from the three-body phase space integral in dimension $D$ \cite{Schwartz:2014sze}.} 
It is easy to see that the infrared
divergences in the  decay amplitude with virtual corrections in Eq.~(\ref{eq:decay_width_virtual}) can be canceled by the real gluon
corrections from three-body radiation in Eq.~(\ref{3-body-decay}), if we sum the one-loop corrections for the two-body decay and three-body decay processes. In this way, we obtain  a finite total   decay width at the $\alpha_s$ order:
\begin{equation}
\Gamma=\Gamma^{(0)}+\Gamma_{\mathrm{virtual}}^{(1)}+\Gamma_{\mathrm{real}}^{(1)}=\Gamma^{(0)}\bigg[1+\dfrac{\alpha_{s}(\mu)}{3\pi}\bigg(-\dfrac{4}{3}\pi^{2}+5\bigg)\bigg]{+\mathcal{O}(\alpha_{s}^{2})},
\label{eq:decay_width_NLO}
\end{equation}
which is in  agreement with the result presented in Ref.~\cite{Jezabek:1988iv}.  

\subsection{Gluon energy cutoff}

In the calculations of the previous section, we included
the contributions from three-body radiation corrections to cancel
the infrared divergences, which introduces part of the three-body corrections. 
In hadron colliders, there are too many jets from QCD background. Experimentally the signal background ratio for two jets final state is lower than the single jet final state. In any detector, there is an energy cut for a jet. 
Thus in our case, if the gluon jet energy is below 
the detector's gluon energy threshold, experiments treat this event as two body decay, even if theoretically it is three body decay. To compare with experimental results, it is necessary
to subtract the contributions from real gluons with energies above the detector's gluon energy threshold. 

In the semi-inclusive decay of the top quark, other heavy-to-light currents are suppressed by the CKM matrix element compared to $ t \to b $ current, and collider experiments have better resolution for \( b \)-jets than that for light-quark jets. This makes the corresponding processes the most likely to be experimentally observed. Therefore, we further consider an energy cutoff \cite{Quintero:2014lqa} in the \( t \to b\pi^{+} \) and \( t \to bD_{s}^{+} \) processes to subtract the three-body effects above the detector threshold, allowing for a direct comparison with collider experiment results. While the energy cutoff can eliminate soft divergences, collinear divergences still exist. To address this issue, we can simply retain the pole mass of the bottom quark as a collinear regulator to avoid collinear divergences,
\begin{equation}
\begin{aligned}
\Gamma_{\mathrm{real}}^{(1),\mathrm{cut}} 
& =\frac{1}{2m_{t}}\,\int \mathrm{d}\Pi \,\frac{1}{3}\sum_{\mathrm{color}}\frac{1}{2}\sum_{\mathrm{spin}}\sum_{\mathrm{\epsilon}}\,|\mathcal{A}_{\mathrm{real}}|^2,
\label{eq:decay_width_cut} 
\end{aligned}
\end{equation}
where $\epsilon$ is the polarization of gluon. 
The decay amplitude $\mathcal{A}_{\mathrm{real}}$ can simply read from Fig. \ref{fig:real}.
For discrete the three body decay from the dominant two body decay as in \cite{Quintero:2014lqa}, we take a gluon energy cut $\delta$ in the three body phase space integral to get the numerical result of three body decay. We chose dimensionless variables of $b$-jet as $x_{b}$, gluon jet as $x_{g}$, pseudo-scalar meson as $x_{P}$ \cite{Schwartz:2014sze}
\begin{equation}
    x_{b}=\dfrac{2k\cdot p_{t}}{m_{t}^{2}}=\dfrac{2E_{b}}{m_{t}}\,,\quad x_{P}=\dfrac{2p\cdot p_{t}}{m_{t}^{2}}=\dfrac{2E_{P}}{m_{t}}\,,\quad x_{g}=\dfrac{2l\cdot p_{t}}{m_{t}^{2}}=\dfrac{2E_{g}}{m_{t}}\,,
\end{equation}
where $l$ is the momentum of gluon jet, $E_{b}$, $E_{g}$ and $E_{P}$ are the energy of the $b$-jet, gluon jet and pseudo-scalar meson respectively. 
Since we chose $\delta$ as the cut of gluon jet, so the minimum of the gluon energy is $E_{g,min}=\delta$, and the range of these variables are
\begin{equation}
\begin{aligned}
    x_{g}&:\dfrac{2\delta}{m_{t}}\sim\dfrac{m_{t}^{2}-m_{b}^{2}}{m_{t}^{2}} ,\\
    x_{b}&:\dfrac{2m_{b}}{m_{t}}\sim\dfrac{m_{t}^{2}+m_{b}^{2}}{m_{t}^{2}} ,\\
    x_{P}&:0\sim\dfrac{m_{t}^{2}-2\delta m_{t}-m_{b}^{2}}{m_{t}(m_t - 2\delta)} .\\
\end{aligned}    
\end{equation}
After transforming the three-body phase space integral in Eq.~(\ref{eq:decay_width_cut}) into an integral over momentum fractions, we can compute the decay width of real gluon radiation above the threshold \( \Gamma_{\mathrm{real}}^{(1),\mathrm{cut}} \). Taking into account the results from Eq.~(\ref{eq:decay_width_NLO}) and (\ref{eq:decay_width_cut}), we finally obtain the decay width with the energy cutoff:
\begin{equation}
\label{eq:decay_widthen_totally}
\Gamma_{t \to b P^+} = \Gamma-\Gamma_{\mathrm{real}}^{(1),\mathrm{cut}}+\mathcal{O}(\alpha_{s}^{2})\,.
\end{equation}

\section{Numerical results}

\begin{table}
\caption{Input parameters in the numerical calculations. \label{Tab:input_parameters}} 
\begin{spacing}{1.9} 
\centering{}%
\begin{tabular}{ll|ll} 
\hline  \hline  
$\quad f_{\pi}=130.2\pm0.8\,\mathrm{MeV}\quad$ &\cite{FlavourLatticeAveragingGroupFLAG:2024oxs}\quad & $\quad f_{K}=155.7\pm0.7\,\mathrm{MeV}\quad$ & \cite{FlavourLatticeAveragingGroupFLAG:2024oxs}$\quad$\\ 
$\quad f_{D}=212.0\pm0.7\,\mathrm{MeV}\quad$ &\cite{FlavourLatticeAveragingGroupFLAG:2024oxs}\quad & $\quad f_{D_{s}}=249.9\pm 0.5\,\mathrm{MeV}\quad$ & \cite{FlavourLatticeAveragingGroupFLAG:2024oxs}$\quad$\\ 
$\quad f_{B}=190.0\pm 1.30\,\mathrm{MeV}\quad$ & \cite{FlavourLatticeAveragingGroupFLAG:2024oxs}\quad & $\quad f_{B_{c}}=434.0\pm 15.0\,\mathrm{MeV}\quad$ & \cite{Colquhoun:2015oha}$\quad$\\ 
$\quad G_{F}=1.1663788\times10^{-5}\,\mathrm{GeV}^{-2}\,\quad$ & \cite{ParticleDataGroup:2024cfk}$\quad$ & $\quad\Gamma_{t}=1.42_{-0.15}^{+0.19}\,\mathrm{GeV}$ & \cite{ParticleDataGroup:2024cfk}\quad \\ 
$\quad m_{t}=172.57\pm0.29\,\mathrm{GeV} \quad$  &\cite{ParticleDataGroup:2024cfk} & 
$\quad m_{b}=4.78\pm0.06 \,\mathrm{GeV} \quad$  &\cite{ParticleDataGroup:2024cfk} \\
$\quad |V_{ud}|=0.97367\pm0.00032$& \cite{ParticleDataGroup:2024cfk}$\quad$ & $\quad |V_{us}|=0.22431\pm0.00085$ & \cite{ParticleDataGroup:2024cfk}$\quad$\\ 
$\quad |V_{ub}|=(3.82\pm0.20)\times10^{-3}$& \cite{ParticleDataGroup:2024cfk}$\quad$ & $\quad |V_{cd}|=0.221\pm0.004$ & \cite{ParticleDataGroup:2024cfk}$\quad$\\ 
$\quad |V_{cs}|=0.975\pm0.006$& \cite{ParticleDataGroup:2024cfk}$\quad$ & $\quad |V_{cb}|=(41.1\pm1.2)\times10^{-3}$ & \cite{ParticleDataGroup:2024cfk}$\quad$\\ 
$\quad |V_{td}|=(8.6\pm0.2)\times10^{-3}$& \cite{ParticleDataGroup:2024cfk}$\quad$ & $\quad |V_{ts}|=(41.5\pm0.9)\times10^{-3}$ & \cite{ParticleDataGroup:2024cfk}$\quad$\\ 
$\quad |V_{tb}|=1.010\pm0.027$& \cite{ParticleDataGroup:2024cfk}$\quad$ & \quad  & \quad\\
\hline  \hline  
\end{tabular} 
\end{spacing} 
\label{input}
\end{table}

In the calculation of the previous section, we presented the decay
width for $t\rightarrow qP^{+}$ up to $\mathcal{O}(\alpha_{s})$.
To compute the numerical results for the branching ratios, we need
to introduce input parameters, which are detailed in  Table~\ref{input}. The strong coupling constant is chosen as $\alpha_{s}(\hat{m}_{t})=0.1090 \pm 0.0014$ \cite{ParticleDataGroup:2024cfk}. According to the Eq.~(\ref{eq:decay_width_LO}) and (\ref{eq:decay_width_NLO}), the numerical results for the decay width at LO and NLO precision are shown in Table~\ref{Tab:barnching_fraction}. The NLO corrections contribute
approximately 10\% compared to LO one.
Some of the decay channels have been calculated at LO in Ref.~\cite{Beneke:2000hk,dEnterria:2023wjq}. We list their results\footnote{The authors of Ref.~\cite{dEnterria:2023wjq} have corrected the 1/9 factor discrepancy recently and updated their data, establishing consistency with our results.} also in Table~\ref{Tab:barnching_fraction}. There exists nearly an order-of-magnitude discrepancy between the results in \cite{Beneke:2000hk} and ours, originating from a difference by a factor of 1/9 in Eq.~(\ref{eq:decay_width_LO}).

Among all the decay channels,  $t\rightarrow b\pi^{+}$ and $t\rightarrow bD_{s}^{+}$ processes have the largest branching ratios. 
 Recent experiments have indicated that the decay processes of the
$D_{s}$ meson series can be measured at the LHC using the ATLAS \cite{ATLAS:2015igt}
and CMS \cite{Mariani:2021bvo} detectors. At the high luminosity LHC in the near future,
the experiments with the highest top quark production rates are the
ATLAS and CMS collaborations.  The processes $t\rightarrow b\pi^{+}$ and $t\rightarrow bD_{s}^{+}$ are
the most likely to be measured in near future experiments. Other decay processes 
are possible to be detected in future collider SPPC \cite{thecepcstudygroup2018cepc,Tang:2022fzs} and
FCC-hh \cite{FCC:2018vvp} with very high energy and high luminosity \cite{dEnterria:2023wjq}. 

\begin{table}[H]
\caption{Branching fractions of top quark semi-inclusive decay at leading order (LO) and next-to-leading order precision (NLO), comparing with previous results. \label{Tab:barnching_fraction}}
\centering{}%
	\begin{spacing}{2}
\begin{tabular}{ccccc}
\hline 
\textbf{Branching fraction} & \textbf{LO (This work)} & \textbf{NLO (This work)} & \textbf{LO}  \cite{Beneke:2000hk} & \textbf{LO} \cite{dEnterria:2023wjq}\\
\hline 
\hline 
$\mathrm{Br}(t\rightarrow b\pi^{+})$ & $1.60_{-0.23}^{+0.19}\times10^{-7}$ & $1.45_{-0.21}^{+0.17}\times10^{-7}$ & $4.0\times10^{-8}$ & $1.6\times10^{-7}$\\
\hline 
$\mathrm{Br}(t\rightarrow bK^{+})$ & $1.22_{-0.18}^{+0.14}\times10^{-8}$ & $1.10_{-0.16}^{+0.13}\times10^{-8}$ & -- & $1.2\times10^{-8}$\\
\hline 
$\mathrm{Br}(t\rightarrow bD^{+})$ & $2.20_{-0.33}^{+0.27}\times10^{-8}$ & $1.99_{-0.30}^{+0.25}\times10^{-8}$ & -- & $2.3\times10^{-8}$\\
\hline 
$\mathrm{Br}(t\rightarrow bD_{s}^{+})$ & $5.93_{-0.86}^{+0.71}\times10^{-7}$ & $5.37_{-0.78}^{+0.64}\times10^{-7}$ & $2.0\times10^{-7}$ & $6.0\times10^{-7}$ \\
\hline 
$\mathrm{Br}(t\rightarrow bB^{+})$ & $5.26_{-0.94}^{+0.84}\times10^{-12}$ & $4.77_{-0.85}^{+0.76}\times10^{-12}$ & -- & $5.3\times10^{-12}$\\
\hline 
$\mathrm{Br}(t\rightarrow bB_{c}^{+})$ & $3.18_{-0.54}^{+0.47}\times10^{-9}$ & $2.88_{-0.49}^{+0.43}\times10^{-9}$ & -- & $3.1\times10^{-9}$\\
\hline 
$\mathrm{Br}(t\rightarrow s\pi^{+})$ & $2.71_{-0.38}^{+0.31}\times10^{-10}$ & $2.46_{-0.35}^{+0.28}\times10^{-10}$ & -- & --\\
\hline 
$\mathrm{Br}(t\rightarrow sK^{+})$ & $2.06_{-0.29}^{+0.24}\times10^{-11}$ & $1.86_{-0.26}^{+0.21}\times10^{-11}$ & -- & --\\
\hline 
$\mathrm{Br}(t\rightarrow sD^{+})$ & $3.70_{-0.54}^{+0.44}\times10^{-11}$ & $3.35_{-0.49}^{+0.40}\times10^{-11}$ & -- & --\\
\hline 
$\mathrm{Br}(t\rightarrow sD_{s}^{+})$ & $1.00_{-0.14}^{+0.12}\times10^{-9}$ & $9.07_{-1.28}^{+1.04}\times10^{-10}$ & -- & --\\
\hline 
$\mathrm{Br}(t\rightarrow sB^{+})$ & $8.89_{-1.56}^{+1.38}\times10^{-15}$ & $8.05_{-1.42}^{+1.25}\times10^{-15}$ & -- & --\\
\hline 
$\mathrm{Br}(t\rightarrow sB_{c}^{+})$ & $5.37_{-0.90}^{+0.78}\times10^{-12}$ & $4.86_{-0.81}^{+0.71}\times10^{-12}$ & -- & --\\
\hline 
$\mathrm{Br}(t\rightarrow d\pi^{+})$ & $1.16_{-0.17}^{+0.14}\times10^{-11}$ & $1.05_{-0.15}^{+0.12}\times10^{-11}$ & -- & --\\
\hline 
$\mathrm{Br}(t\rightarrow dK^{+})$ & $8.84_{-1.26}^{+1.03}\times10^{-13}$ & $8.00_{-1.14}^{+0.93}\times10^{-13}$ & -- & --\\
\hline 
$\mathrm{Br}(t\rightarrow dD^{+})$ & $1.59_{-0.23}^{+0.19}\times10^{-12}$ & $1.44_{-0.21}^{+0.17}\times10^{-12}$ & -- & --\\
\hline 
$\mathrm{Br}(t\rightarrow dD_{s}^{+})$ & $4.30_{-0.61}^{+0.50}\times10^{-11}$ & $3.90_{-0.55}^{+0.45}\times10^{-11}$ & -- & --\\
\hline 
$\mathrm{Br}(t\rightarrow dB^{+})$ & $3.82_{-0.67}^{+0.60}\times10^{-16}$ & $3.46_{-0.61}^{+0.54}\times10^{-16}$ & -- & --\\
\hline 
$\mathrm{Br}(t\rightarrow dB_{c}^{+})$ & $2.31_{-0.39}^{+0.34}\times10^{-13}$ & $2.09_{-0.35}^{+0.31}\times10^{-13}$ & -- & --\\
\hline 
\end{tabular}
\end{spacing}
\end{table}

The energy threshold for jets are usually detector dependent \cite{CMS:2020cmk,ATLAS:2024xna}. In our calculation,
 we set the gluon energy cutoff thresholds to $\delta$=20, 25, 30\,GeV, respectively. The branching ratios of  $t\rightarrow b\pi^{+}$ and $t\rightarrow bD_{s}^{+}$
processes with such different energy cuts of $\Gamma_{t\rightarrow bP^{+}}$ in Eq.~(\ref{eq:decay_widthen_totally})  are presented
 in Table~\ref{Tab:branching_fraction_cut}.
To avoid collinear divergences
in our calculation, we retain the mass of
the bottom quark \cite{Quintero:2014lqa} in the $t\rightarrow b\pi^{+}$ and $t\rightarrow bD_{s}^{+}$
processes. From Table~\ref{Tab:branching_fraction_cut}, one can see that the branching ratios of experimental detectable two body decays have around 20\% differences for different energy cut of gluon jet. 

For the future collider HL-LHC, running at a center-of-mass energy of \(\sqrt{s} = 14\,\mathrm{TeV}\), the total integrated luminosity is expected to be \(3\,\mathrm{ab^{-1}}\)~\cite{Apollinari:2015wtw}, with the top quark pair production cross section approximately \(\sigma_{t\bar{t}}^{\mathrm{HL-LHC}} = 1\,\mathrm{nb}\)~\cite{FCC:2018vvp}. For the future collider FCC-hh, the integrated luminosity is projected to be \(20\,\mathrm{ab^{-1}}\)~\cite{FCC:2018vvp} at energies $\sqrt{s}= 100$ {TeV}, with the top quark pair production cross section around \(\sigma_{t\bar{t}}^{\mathrm{FCC-hh}} = 35\,\mathrm{nb}\) \cite{FCC:2018vvp}. Based on the total production cross section and integrated luminosity for future colliders, the number of generated events can be estimated using the formula provided in Ref.~\cite{dEnterria:2020ygk} :
\begin{equation}
     N_{\mathrm{event}}  =\sigma_{t\bar{t}}\times\mathcal{L}_{int}\times\mathrm{Br}(t\rightarrow qP^{+}),
\end{equation}
where $\mathcal{L}_{int}$ is the total the integrated luminosity of future collider. Based on the branching fraction with the cutoff in Table~\ref{Tab:branching_fraction_cut}, it can be seen that the number of events for the \( t \to b\pi^+ \) and \( t \to bD_s^+ \) processes at the HL-LHC is expected to be in the range of \( 10^2 \sim 10^3 \), making their detection possible in the near future. For the future collider FCC-hh, the event yield for these two dominant rare decay channels is projected to be in the range of \( 10^4 \sim 10^5 \), providing a sufficient dataset for precise tests of these processes. These two rare decay processes can be used to test the QCD factorization theorem in top quark semi-inclusive decays, as well as serve as probes for new physics.

\begin{table}
\caption{Branching fractions of $t\rightarrow b\pi^{+}$ and $t\rightarrow bD^{+}_{s}$ decay with gluon energy cut $\delta$ = 20, 25, 30 GeV, respectively. }\label{Tab:branching_fraction_cut}
\centering{}%
\begin{spacing}{2}
\begin{tabular}{cccc}
\hline 
$\bf{Branching\;fraction}$ & $\bf{NLO\, (\delta=20 GeV)}$ & $\bf{NLO\, (\delta=25 GeV)}$ & $\bf{NLO\, (\delta=30 GeV)}$\\
\hline 
\hline 
$\mathrm{Br}(t\rightarrow b\pi^{+})$ & $8.39^{+1.10}_{-1.09}\times10^{-8}$ & $9.89\pm 1.29\times10^{-8}$ & $1.10\pm 0.14\times10^{-7}$\\
\hline 
$\mathrm{Br}(t\rightarrow bD_{s}^{+})$ & $3.10^{+0.41}_{-0.40}\times10^{-7}$ & $3.65\pm 0.48\times10^{-7}$ & $4.05\pm 0.53\times10^{-7}$\\
\hline 
\end{tabular}
\end{spacing}
\end{table}

\begin{table}
\caption{The estimation of the event numbers for the top quark semi-inclusive decay processes at future colliders.} \label{Tab:eventnumber}
\centering{}%
	\begin{spacing}{2}
\begin{tabular}{cccc}
\hline 
   \textbf{Event number}	& $\bf{NLO\, (\delta=20 GeV)}$	& $\bf{NLO\, (\delta=25 GeV)}$	& $\bf{NLO\, (\delta=30 GeV)}$\\
    \hline 
    \hline 
$N_{\mathrm{even}}^{\mathrm{HL-LHC}}(t\rightarrow b\pi^{+})$ &	$2.5\times10^{2}$	& $3.0\times10^{2}$	& $3.3\times10^{2}$\\
    \hline 
$N_{\mathrm{even}}^{\mathrm{HL-LHC}}(t\rightarrow bD_{s}^{+})$ &	$9.3\times10^{2}$  &	$1.1\times10^{3}$ &	$1.2\times10^{3}$\\
    \hline 
$N_{\mathrm{even}}^{\mathrm{FCC-hh}}(t\rightarrow b\pi^{+})$ &	$5.9\times10^{4}$	& $6.9\times10^{4}$ &	$7.7\times10^{4}$\\
    \hline 
$N_{\mathrm{even}}^{\mathrm{FCC-hh}}(t\rightarrow bD_{s}^{+})$ &	$2.2\times10^{5}$	& $2.6\times10^{5}$ &	$2.8\times10^{5}$\\
\hline 
\end{tabular}
\end{spacing}
\end{table}

\section{Conclusion}

In this work, we have calculated the NLO corrections to the top quark semi-inclusive decay \( t \rightarrow qP^{+} \). Within the framework of QCD factorization, the leading power matrix element is factorized into inclusive and exclusive components up to NLO. The ultraviolet and infrared divergences at NLO are canceled in the computation of the leading power decay width. We present numerical results for the branching fractions of the two body semi-inclusive processes, including the corresponding real gluon emission three body decays. The branching ratios for these processes span a wide range, from \( 10^{-16} \) to \( 10^{-7} \), due to quite different CKM matrix elements. 

For the  $t\rightarrow b\pi^{+}$ and $t\rightarrow bD_{s}^{+}$
processes, with the largest CKM matrix elements, we provide their branching ratio dependence on the cutoff gluon energy, anticipating that they could be measured at the near future experiments. These results  provide a chance to test the factorization theorem in semi-inclusive processes, which could also serve as a probe for new physics.

\section*{Acknowledgment}

{We are grateful to Dong-Hao Li for helpful discussions.} The work is supported  in part by the National Key Research and Development Program of China
(2023YFA1606000) and by the National Natural Science Foundation of China under Grant No.~12447185, 12275277 and 12435004. 

\newpage

\bibliographystyle{apsrev4-2}
\addcontentsline{toc}{section}{\refname}
\bibliography{Reference}

\end{document}